%% file: main.tex
\documentclass[%
reprint,10pt,
superscriptaddress,
amsmath,amssymb,
aps,prl
]{revtex4-1}
\usepackage[margin=1in]{geometry}
\usepackage[utf8]{inputenc}
\usepackage[hidelinks]{hyperref}
\usepackage{graphicx}
\usepackage{amsmath,amssymb}

\usepackage{wrapfig,lipsum}
\usepackage{float}

\usepackage{appendix}
\usepackage{titletoc}

\usepackage{bbm}

\titlecontents{section}
  [0pt] 
  {\vspace{0.5em}\bfseries} 
  {\thecontentslabel\enspace} 
  {} 
  {\titlerule*[0.5pc]{}\contentspage} 

\titlecontents{subsection}
  [1.5em] 
  {\normalfont} 
  {\thecontentslabel\enspace} 
  {} 
  {\titlerule*[0.5pc]{}\contentspage} 

\newcommand\DoToC{%
  \startcontents
  \printcontents{}{1}{\vskip5pt\hrule\vskip1pt}
  \vskip3pt\hrule\vskip5pt
}

\usepackage{hyperref}
\usepackage{xcolor}

\begin{document}
\title{Glass-like Caging with Random Planes}
\author{Gilles Bonnet}
\affiliation{Bernoulli Institute, University of Groningen, Groningen, Netherlands}
\affiliation{CogniGron, University of Groningen, Groningen, Netherlands.}
\author{Patrick Charbonneau}
\affiliation{Department of Chemistry, Duke University, Durham, North Carolina 27708}
\affiliation{Department of Physics, Duke University, Durham, North Carolina 27708}
\author{Giampaolo Folena}
\email{giampaolofolena@gmail.com}
\affiliation{Department of Chemistry, Duke University, Durham, North Carolina 27708}

\input{maintext}

\bibliography{biblio.bib,intro.bib}
\pagebreak
\onecolumngrid
\pagebreak
\noindent{\centering\large\bfseries Supplementary Material\par}
\begin{appendices}
\DoToC
\input{auxiliaries}
\end{appendices}
\end{document}

%% file: maintext.tex





\begin{abstract}
The richness of the mean-field solution of simple glasses leaves many of its features challenging to interpret. A minimal model that illuminates glass physics the same way the random energy model clarifies spin glass behavior would therefore be beneficial. Here, we propose such a real-space model that is amenable to infinite-dimensional $d\rightarrow\infty$ analysis and is exactly solvable in finite $d$ in some regimes. By joining analysis with numerical simulations, we uncover geometrical signatures of the dynamical and jamming transitions and obtain insight into the origin of activated processes. Translating these findings into the context of standard glass formers further reveals the role played by non-convexity in the emergence of Gardner and jamming physics.
\end{abstract}

\maketitle 

\paragraph{Introduction --}
The prototypical random energy model (REM)~\cite{Derrida1980,Derrida1981} plays a key pedagogical role in the study of disordered systems~\cite{MPV1986,nishimori2001statistical,mezard2009information,sethna2021statistical}. Comparing probabilistic and theoretical physics descriptions of the model notably provides key insight into the nature of the replica symmetry breaking (RSB) formalism. The recent formulation of a first-principle theory of simple glasses by porting the RSB approach to real-space systems in the mean-field infinite-dimensional limit $d\rightarrow\infty$~\cite{book_2020} could benefit from a REM-like reference. The richness of the theoretical description indeed at times obfuscates its interpretation. The role of geometry in the dynamical arrest, the nature of the rare escape (instantonic) trajectories, and the robustness of jamming criticality are but a few of its stupefying features.

Simplifying a model to effectively investigate its essence, however, is easier said than done. In the case of simple glasses even drastic truncations, such as removing all multi-body correlations in the Mari-Kurchan model~\cite{Mari2009,Mari2011} or immobilizing all but one particle in the random Lorentz gas (RLG)~\cite{jin2015dimensional}, result in mean-field descriptions that are no simpler than the original one~\cite{bir2021a,bir2021b,Charb2021,bir2022}. What minimal description could be proposed to achieve REM-like grasp of the system? 

An oft-used simplifying strategy for real-space systems is to construct a convex cell version of the model. The cell, which corresponds to the (free volume) cavity explored by an individual sphere while fixing all others, can then be thermodynamically analyzed~\cite{hoover1972exact,Speedy1981CavitiesAF}.  That approach has a long track record, from the Lennard-Jones--Devonshire cell model of liquids~\cite{lennard1937critical,lennard1938critical}, to cell models of both ordered \cite{buehler1951,wood1952note,salsburg1962equation} and amorphous  \cite{ziman1979models,hoover1979exact,sastry1997,sastry1998free} solids, albeit with mixed success. While it has since been largely discarded for the former~\cite{rowlinson2015}, it remains squarely in use for the latter~\cite{royall2023colloidal}, efficiently capturing, for instance, the scaling of the pressure divergence upon approaching close packing~\cite{donev2005,kamien2007liu,kamien2007}. 

In this spirit, we here propose to consider a convex cell version of the RLG, the hyperplane-RLG (hRLG), which is singularly amenable to analysis. Its exact solution in the mean-field limit $d\rightarrow\infty$ can be obtained using the RSB method, and both its high- and low-density properties can be related to models of Poisson hyperplane tessellation (PHT) that were recently solved for all $d$ \cite{horrmann_volume_2014,kabluchko_angles_2020,kabluchko_expected_2020}. By combining these two descriptions, key insights into the role of non-convexity in glass formation are obtained. Moreover, by defining an analogue of jamming, we find that isostaticity can emerge without the need for a Gardner transition. 

\begin{figure}[t]
    \centering
		\includegraphics[width=0.7\columnwidth]{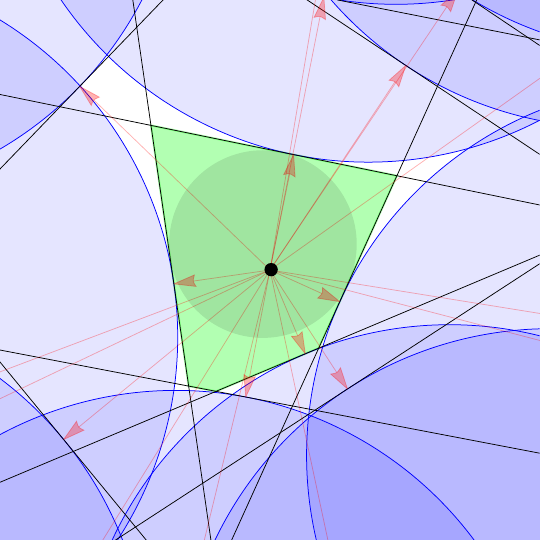}\\
		\caption{Sample hRLG construction in $d=2$ for a point tracer initially at $x_0$ (black dot). The hRLG cell (green polygon) is obtained by convexifying the RLG cage, which is the space complementary to the obstacles (blue disks). The process is equivalent to randomly sampling hyperplanes (black lines) normal to isotropically distributed vectors of length $H$ taken from $P(H)$ (red arrows). 
        The inscribed sphere of the hRLG cell (dark green circle) is unique and isostatic. It corresponds to the inherent structure obtained by growing the tracer.}  
  \label{fig:Cell}
\end{figure}

\paragraph{Model and exact results--}
To rigorously define the hRLG model, first recall that the RLG consists of non-interacting spherical obstacles of radius $l$ distributed uniformly at random in $\mathbb{R}^d$ with number density (or intensity) $\rho=N/V$. 
Considering a random origin $\mathbf{x}_0$ such that it lies inside a RLG cage (void between obstacles),
there exists a one-to-one map from obstacles to tangent planes (see Fig.~\ref{fig:Cell}). (In the limit $\rho\rightarrow\infty$ the RLG is congruent to the hRLG, because small cages are then bounded by nearly flat surfaces.) The probability distribution function of the distance $H$  between a hyperplane and that origin is then
\begin{equation}\label{eq:Ph}
    P(H)\mathrm{d}H\propto
    (H+l)^{d-1}\mathrm{d}H \qquad \text{for} \quad  H>0  \ .
\end{equation} 
Although $P(H)$ diverges with $H$, the number of planes that actually delimit a cage is finite. We can therefore define a cutoff $H_\mathrm{max}$ to the distribution. In numerical simulations, we fix the total number $M$ of hyperplanes sampled, such that $M = \alpha_M \#$(facets of a typical cell) with $\alpha_\mathrm{M}\gg1$ but not too large so as to limit computational costs. (In practice, $\alpha_M=200$ is found to be a good compromise.) The resulting cumulative probability (setting the radius to unity, $l=1$) is then
\begin{equation}\label{eq:C_RLG}
    \mathcal{C}(H) =
    \frac{d\widehat{\varphi}}{M} [(1+H)^d-1 ]  \qquad \text{for} \quad 0<H<H_\mathrm{max}
\end{equation} 
and $\mathcal{C}(H>H_\mathrm{max}) = 1$ with $H_\mathrm{max} = (\frac{M}{d\widehat{\varphi}}+1)^{1/d}-1$, using the dimensionally rescaled volume fraction $\widehat{\varphi}=\rho V_d/d$ with $V_d$ the volume of the $d$-dimensional ball of unit radius.

The cage formed by random planes around the origin defines the zero cell of a PHT process. In each instance, that cell varies in shape and size, but its average geometric features are well-defined. Recent results from stochastic geometry evaluate certain of these quantities for PHT with homogeneous cumulative probability 
\begin{equation}\label{eq:C_hom}
    \tilde{C}(H)=
    \frac{d^2\widehat{\varphi}}{rM} H^{r} \qquad \text{for} \quad 0<H<(\frac{rM}{d^2\widehat{\varphi}})^{1/r} ,
\end{equation} 
where $r\in(0,\infty)$ is the distance exponent \cite{math}. In particular, for $r=1,d$ exact finite-$d$ results have been obtained for  the average number of $k$-faces~\cite{kabluchko_angles_2020,kabluchko_expected_2020} as well as the average cell volume and its fluctuations~\cite{horrmann_volume_2014}. 

 \begin{figure}[t]
    \centering
	\includegraphics[width=\columnwidth]{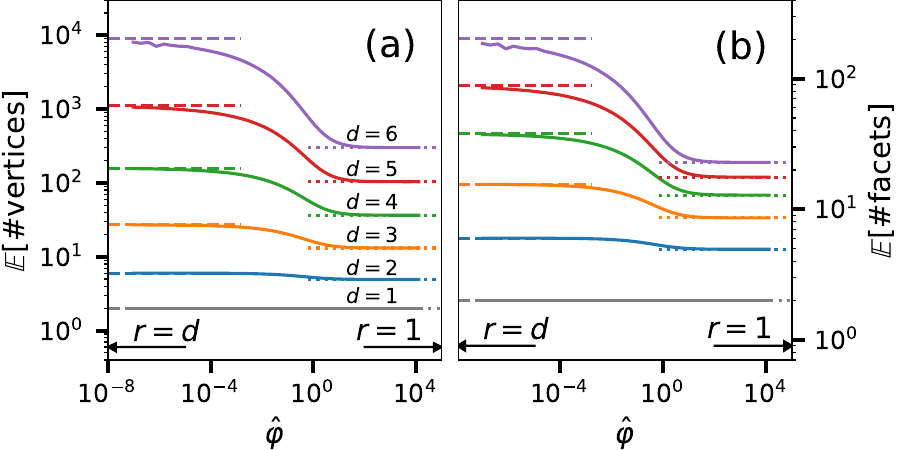}
		\caption{Comparison of hRLG simulation results in  $d=1\ldots6$ (full lines) with the zero cell results of the PHT with $r=1$~\cite{kabluchko2020expected} (dotted lines) and $r=d$~\cite[Thm.~3.3]{kabluchko2021recursive} (dashed lines), to which the model corresponds in the limits $\widehat{\varphi}\to\infty$ and $\widehat{\varphi}\to 0$, respectively. The average number of (a) vertices (0-faces) and (b) facets ($(d-1)$-faces) smoothly interpolates between these two limits, but upon increasing $d$ the crossover becomes more pronounced. Scaling analysis suggests that the discontinuity converges to $\widehat{\varphi}=0$ in the limit $d\rightarrow\infty$~\cite{SM}.}
  \label{fig:Limits}
\end{figure}

Direct comparison of Eqs.~\eqref{eq:C_RLG} and \eqref{eq:C_hom} gives that two PHT cases are recovered as limits of the hRLG model: $r=1$ is attained in the small cell limit $\widehat{\varphi}\to\infty$ (or equivalently $H_\mathrm{max}\to 0$); and $r=d$ is attained in the Poisson-Voronoi tessellation limit $\widehat{\varphi}\to 0$ (or $H_\mathrm{max}\to \infty$)  \cite{horrmann_volume_2014}. Note that other $P(H)$ distributions could be recovered by generalizing the hRLG model to different potentials for sampling obstacles around the central one or by softening the particle-obstacle interaction~\cite{Ph,SM}.

\paragraph{Results --}
Figure~\ref{fig:Limits} shows the robust agreement between PHT results for $r=1$ and $r=d$ with numerical hRLG tessellations obtained using the qhull~\cite{barber1996quickhull} and scipy~\cite{SciPy2020} packages.  Interestingly, the average number of vertices and facets in the two limiting cases of $r$ scale differently with $d$, i.e., $\lim_{d\to\infty}\mathbb{E}[\#\text{vertices},\#\text{facets}] \propto e^{d\log(\pi)}$ for $r=1$ and  $\propto e^{d\log(2\pi d)/2}$ for $r=d$ \cite{Horrmann2015}. In order for the super-exponential ($r=d$) and the exponential ($r=1$) scaling to match, the finite-$d$ crossover in  $\widehat{\varphi}$ grows increasingly sharp with $d$. This crossover suggests the emergence of a marked change in the geometry of the typical polytope~\cite{bonnet2022}, which we conjecture turns into a sharp geometrical transition at $\widehat{\varphi}=0$ in the limit $d\to\infty$~\cite{SM}. 
Because this transition involves the cell \emph{surface} and not its volume, however, it is not thermodynamic in nature, and therefore a standard Monte Carlo or Newtonian dynamics exploration of the cell would be blind to this particular feature. 

\begin{figure}[t]
    \centering
		\includegraphics[width=0.99\columnwidth]{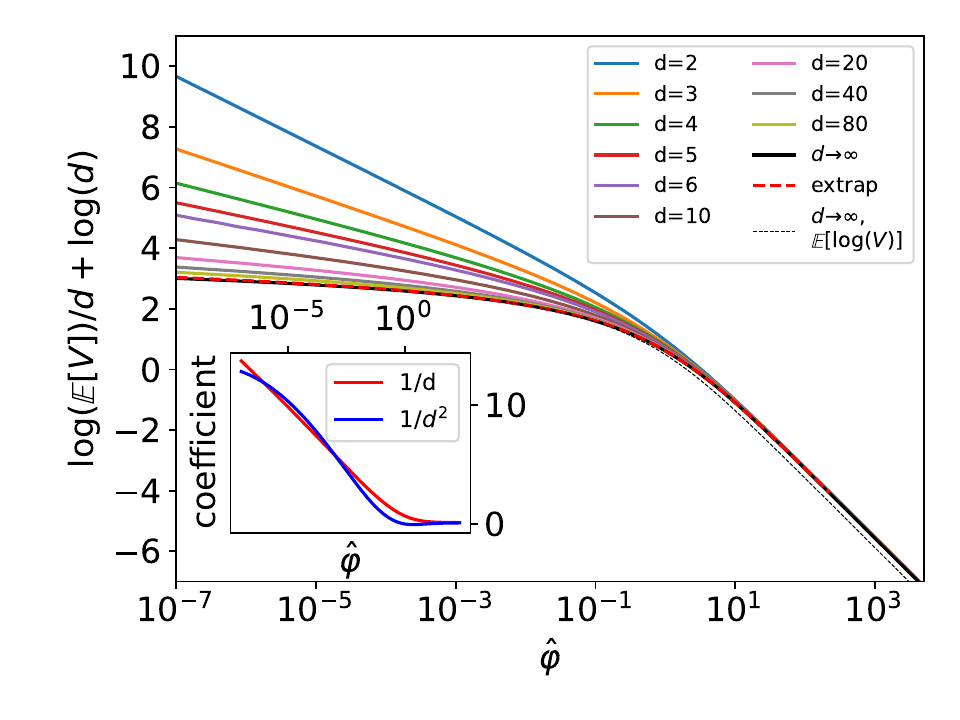}\\
		\caption{Average cell volume in the hRLG model from the RSB method for the annealed (solid black line) and quenched (dashed black line) estimates, as well as exact finite $d$ results for the average cell volume for $d=2\ldots80$ (solid color lines). (Results for $d=1$ are off trend.) A quadratic fit to the finite $d$ results extrapolate to the $d\rightarrow\infty$ prediction with increasing difficulty as $\widehat{\varphi}\rightarrow0$ (red dashed line). (Inset) The linear and quadratic coefficients of these fits grow as density decreases, as a result of the increase in fluctuations.}  \label{fig:AnnVol}
\end{figure}

To see if a clearer thermodynamic signature of this effect could be obtained, we next consider the hRLG model in the mean-field limit of $d\to\infty$, which can be studied analytically  using the RSB method. This approach notably provides the average volume $\mathbb{E}[V]$, the average  of the logarithm of the volume $\mathbb{E}[\log(V)]$, and the variance of the distance from the center of mass of the cage (or the long-time mean squared displacement for a reversible dynamics) $\mathbb{E}[\Delta]=\mathbb{E}[\int_{V_\mathrm{cell}}\!\!\!d\mathbf{x} \; |\mathbf{x}-\mathbf{x}_c|^2/V_\mathrm{cell}]$. 
An annealed calculation provides a closed expression for the average volume~\cite{SM}
\begin{equation}\label{eq:annvol1}
\mathbb{E}[V_\mathrm{cell}] \sim e^{d F[\Delta^*]-d^2\log(d)} 
\end{equation}
with
\begin{equation}\label{eq:annvol2}
\scriptstyle
F[\Delta] = 
\frac{1}{2}\log(\pi e \Delta) +  \frac{\widehat{\varphi}}{2} \left(1-e^{\Delta /4}\left(1+\text{erf}\left(\frac{\sqrt{\Delta }}{2}\right)\right) \right)
\end{equation}
and $\Delta^*=\mathbb{E}[\Delta]$ is the saddle point solution
\begin{equation}\label{eq:annvol3}
    \widehat{\varphi}^{-1} = \frac{\sqrt{\Delta^* }}{2 \sqrt{\pi }}+\frac{1}{4} e^{\Delta^* /4} \Delta^*  \left( 1 + \text{erf}\left(\frac{\sqrt{\Delta^* }}{2}\right)\right) \ .
\end{equation}
A quenched computation evaluated numerically also gives $\mathbb{E}[\log(V)]$  (see Fig.~\ref{fig:AnnVol}). The two results nearly coincide at all $\widehat{\varphi}$. Upon closer inspection, however, we note that the second converge to the first only at small $\widehat{\varphi}$, meaning that sample-to-sample fluctuations are then suppressed. Another key distinction is that the  rotational symmetry of the annealed computation gives that the center of mass and the origin of the cell coincide, i.e., $\mathbf{x}_0=\mathbf{x}_c$, whereas the quenched computation finds that the center of mass of a typical cell does not coincide with its origin, i.e. $\mathbf{x}_0\neq\mathbf{x}_c$. That effect also vanishes at small $\widehat{\varphi}$ (large cages).

Surprisingly, the results from Eqs.~\eqref{eq:annvol1}--\eqref{eq:annvol3} converge to the $d\to\infty$ PHT predictions only for large $\widehat\varphi$ (where $r=1$). To obtain a clearer understanding of the low-density limit behavior (where $r=d$), we obtain 
finite-$d$ analytical results for the average volume~\cite{SM}, which recapitulate the PHT results at both high and low densities. Figure~\ref{fig:AnnVol} shows that 
at small density the curves spread apart, with the gap steadily diverging as $\widehat{\varphi}\rightarrow0$. Although the finite-density results all clearly extrapolate to the $d\rightarrow\infty$ curve, ever higher order corrections are needed to do so upon approaching $\widehat{\varphi}=0$. The limits $\widehat{\varphi}\to 0$ and $d\to\infty$ therefore do not commute. This phenomenon is consistent with the existence of a dynamical transition at $\widehat{\varphi}=0$ in the mean-field $d\rightarrow\infty$ hRLG. This transition, which gives rise to diverging fluctuations \cite{Fol2022}, is associated with cage/cells formation, in close analogy with the formation of states at infinite temperature in the REM. We further observe that as the limit $d\to\infty$ is approached, the primary distinction between RLG and hRLG lies in the curvature of the spherical obstacles. In essence, the RLG (as $d\to\infty$) is analogous to a random paraboloid model, where the paraboloids are tangent to the hRLG planes. Consequently, introducing any constant curvature component to the obstacles, thus breaking convexity, shifts cell formation to finite $\widehat{\varphi}$.

\begin{figure}[t]
    \centering
		\includegraphics[width=0.99\columnwidth]{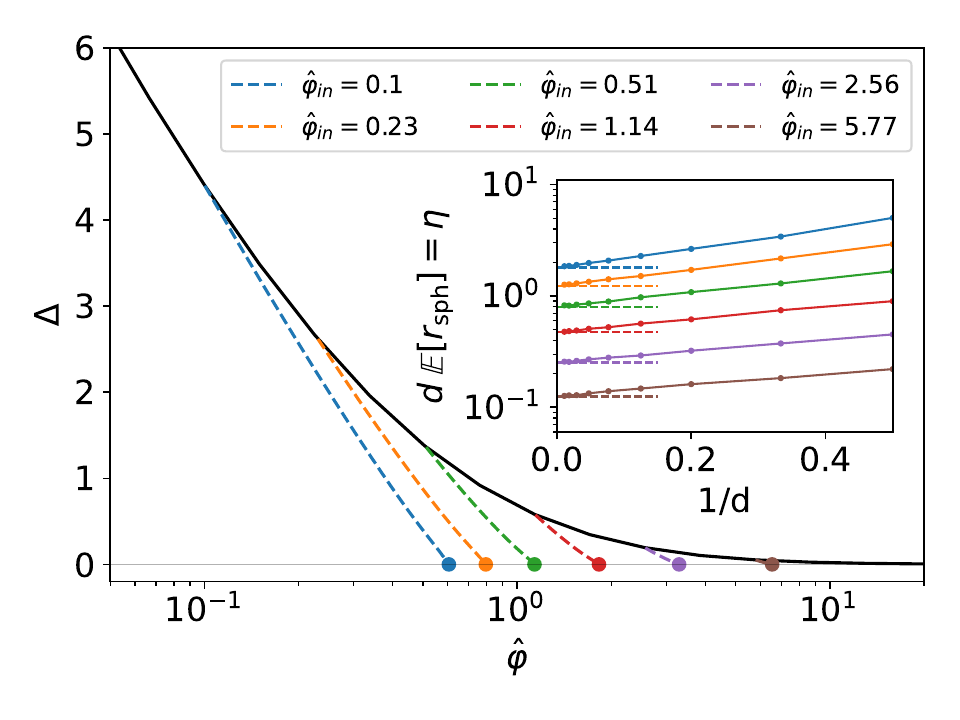}\\
		\caption{Compression (state-following) procedure for the hRLG in the mean-field limit $d\to\infty$  for different initial $\widehat{\varphi}_\mathrm{in}$ (dashed colored lines). Each colored dashed line follows the change in the average second moment $\Delta=2d \mathbb{E}[\frac{\int_{V_\mathrm{cell}} d\mathbf{x} (\mathbf{x}-\mathbf{x}_c)^2}{V_\mathrm{cell}}]$ during a compression from different initial $\widehat{\varphi}_\mathrm{in} = 0.1,0.23,0.51,1.14,2.56,5.77$. Because $\widehat{\varphi}_\mathrm{in}=0$ jams at zero density, the jamming line here extends over the whole $\widehat{\varphi}$ axis. (Inset) Simulation results for in $d=2\ldots89$ (points; lines are guides for the eyes) for the inherent structure density for the hRLG at the same $\widehat{\varphi}_\mathrm{in}$ as in the main panel.  The average insphere radius is expected to scale asymptotically as $\mathbb{E}[r_\mathrm{sph}] = \eta/d$, where $\eta$ can be determined from the mean-field $d\rightarrow\infty$ description (dashed lines)~\cite{SM}.}  \label{fig:SF}
\end{figure}

In order to identify the jamming transition of the hRLG~\cite{vanHecke2010,liu2010,rainone2015,book_2020}, we next consider the equivalent of a compression protocol. In the hRLG, cell compression is akin to growing the tracer until the volume available to its center is but a single point. This process, which is analogous to identifying the inherent structure in the REM and other models, corresponds to determining the inscribed sphere (or insphere) for the cell. The tracer is then expected to be confined within an isostatic simplex composed of $d+1$ hyperplanes~\cite{SM} (see Fig.~\ref{fig:Cell}). In the mean-field $d\rightarrow\infty$ limit, the radius $r_\mathrm{sph}$ and the distance $|\mathbf{x}_\mathrm{sph}|$ from the origin $\mathbf{x}_0$ can be computed using the RSB method and the quasi-equilibrium state-following construction. The approach properly recovers isostaticity and Fig.~\ref{fig:SF} shows the robust agreement between $d\rightarrow\infty$ result and extrapolations from finite-$d$ simulations obtained by linear optimization \cite{SM}. 

Interestingly, the purely convex nature of the hRLG results in an unambiguous (bijective) inherent structure mapping. By contrast, in the limit $d\rightarrow\infty$, a comparable compression procedure for the RLG necessarily undergoes a Gardner transition, beyond which the cage is defined by an ultrametric structure of sub-cages, thus resulting in a large response to small perturbation, i.e., marginality. Although both scenarios result in isostatic jammed configurations, their outcome differs in one key respect. The jammed states of the hRLG have dimensionally-robust \emph{trivial} critical exponents that describe microstructure at jamming~\cite{SM,perceptron}, whereas dimensionally-robust \emph{non-trivial} critical exponents result from Gardner physics~\cite{charb2017,franz2017}.

The close analogy between the RLG and the hRLG nevertheless suggests a practical procedure for determining inherent structures in the RLG. Recall that the cell construction in going from the RLG to the hRLG depends on the choice of origin $\mathbf{x}_{0}$. That cell is therefore only one possible convexification of the RLG. Each $\mathbf{x}_{0}$ can be mapped (by the compression procedure) into a different $\mathbf{x}_\mathrm{sph}$ with corresponding insphere radius $r_\mathrm{sph}$. Any jump in $r_\mathrm{sph}$ by continuously moving $\mathbf{x}_{0}$ signals a change in basin of attraction and corresponding inherent structure. Equivalently one can build an algorithm that, given any initial caged point $\mathbf{x}_\mathrm{in}$ in the RLG, finds the closest inherent structure to it. First $\mathbf{x}_\mathrm{sph}$ is found (by linear optimization) and then the relative $r_\mathrm{sph}$ is locally optimized by moving $\mathbf{x}_{0}$, so as to maximize its growth. The scheme converges on $r_\mathrm{sph}$ with corresponding $\mathbf{x}_\mathrm{sph}$ being the inherent structure for the initial point $\mathbf{x}_\mathrm{in}$. This volume ascent (by analogy with gradient descent) algorithm further allows for the partitioning of the RLG cage in different basins, and therefore the direct study of its jamming criticality. This gedankenexperiment moreover suggests that the ultrametric Gardner phase can be a local geometrical phenomenon that only requires a single cage to be observed, although only in the limit $d\rightarrow\infty$.

\paragraph{Conclusion--}
We have introduced the hRLG model as a real-space analogue of the REM. This convex model of glass-like caging is amenable to direct treatment using the replica method in the limit $d\rightarrow\infty$ as well as to formal analysis for certain observables in all $d$. Bringing together these two perspectives  reveals the existence of a geometrical transition of the shape of the cell surface that corresponds to a zero-density dynamical transition. Pushing the geometric analogy further allows us to formulate an inherent structure determination procedure for the hRLG. Remarkably, the resulting jammed states are isostatic with trivial structural critical exponents. The procedure also informs the formulation of a comparable procedure for the RLG, which should enable the exploration of the single-particle nature of Gardner and jamming physics in that model. 

More generally, the hRLG enriches our understanding of the role of non-convexity (obstacle curvature) in glasses. It is found to be key for having a finite-density dynamical transition in $d\rightarrow\infty$. Because a purely convex model of caging does not permit \emph{any} escape paths,  we also understand activated (instantonic) paths, such as the entropic bottlenecks that allow to sidestep that transition in finite $d$, to be deeply reliant on non-convexity. Put differently, geometrical convexity underlies fast configurational sampling, and hence glassiness is intimately linked to non-convexity. Because Gardner physics (multibasin hierarchy) and non-trivial jamming criticality also emerge thanks to obstacle curvature, cage \textit{flatness} should be considered an indicator of the absence of Gardner physics. Systems of binary mixtures with large size asymmetry or with broad size polydispersity likely belong to this class, but have yet to be carefully examined under this lens. Particle softness, which also causes Gardner avoidance \cite{scalliet2017,book_2020}, might analogously find the hRLG to be a natural reference model system.

Finally, the hRLG suggests a very simple geometrical connection between jamming and equilibrium cages through the inscribed sphere construction. This connection offers a promising step toward resolving the mathematical conundrum of constructing a model for a 'random jammed packing of hard balls'~\cite[p.~240-242]{chiu2013stochastic}.


\paragraph{Acknowledgments--}
We would like to thank Francesco Zamponi for analytical guidance, and GB thanks Zakhar Kabluchko for a useful discussion. This work was supported by a grant from the Simons Foundation (Grant No.~454937). GF also acknowledges support from a postdoctoral fellowship from the Duke Center on Computational Thinking.  Data relevant to this work have been archived and can be accessed at the Duke Digital Repository \url{https://doi.org/10.7924/XXXXXXX}.

%% file: auxiliaries.tex


\section{\lowercase{h}RLG simulations}

Recall that the hRLG cell is convexified from a RLG cage built around a tracer at $x_0$. To generate a cell (a random convex polytope) we use a planting scheme~\cite{Biroli2022}. We sample $M$ values of $H_i$ from the radial distribution of hyperplanes Eq.~\eqref{eq:Ph},  and associate to each a random unit vector $\mathbf{V}_i$, thus defining the distance of the hyperplane from the origin and its normal.  (Although $M$ hyperplanes are generated, only a small fraction actually define the cell, i.e., $\mathbb{E}[\#\mathrm{facets}]\ll M$.)
The cell is then given by the intersection of the $M$ half spaces
\begin{equation}\label{eq:half}
   \mathbf{V}_i  \cdot  \mathbf{x} \leq H_i \quad \text{for} \quad i = 1\dots M \ .
\end{equation}
Different algorithms are used to determine: A.~the number of facets and vertices of a cell; B.~a cell volume and its second moment; C.~the position and size of the largest inscribed sphere in a cell. 

\subsection{Convex hull of the dual cell: facets and vertices} 
The simplest way to identify the hyperplanes ($\approx \mathbb{E}[\#\mathrm{facets}]$) that contribute to the  cell is to evaluate all intersection points between hyperplanes by solving all possible linear equations involving $d$ hyperplanes $\mathbf{V}_{\pi_{i}} \cdot \mathbf{x} =  H_{\pi_{i}}$ for $\quad i = 1\dots d$, with $\mathrm{\pi_{i}}_{\neq i}$ is one of the $\binom{M}{d}$ collections of $d$ hyperplanes between $M$. The resulting points which respect Eq.~\eqref{eq:half} are the vertices of the cell from which the facets can be deduced. Given the factorial in $M$ computational cost of this approach, however, it is impractical.

A more efficient approach consists of solving the convex hull of the dual structure, obtained by matching hyperplanes to points~\cite{preparata2012computational}. The intersection of the corresponding hyperspaces (cell) is then mapped to the convex hull of dual points. This algorithm, which is implemented in the qhull package~\cite{barber1996quickhull}, returns the set of facets and vertices of the resulting convex polytope (cell). Although the computational cost of this approach scales markedly more favorably than the brute-force scheme, $M\log M$ v $M!$, its reach remains limited. Because $M\ll\# \mathrm{facets}$, which scales at least exponentially with $d$ (see main text)-- algorithmic complexity also scales at least exponentially with dimension $d$. (In practice, prior studies have reached at most $d=9$~\cite{YiHu2021}.) Random polytopes have an inherent complexity given by their large number of facets/vertices (exponential for $r=1$ and factorial for $r=d$), thus lower bounding algorithmic complexity. The convex hull algorithm --by cleverly selecting the relevant facets/vertices-- saturates that bound. As a result, even in the (worst) factorial case, the prefactor is significantly smaller than for the brute-force algorithm.

\begin{figure}[t]
    \centering
		\includegraphics[width=0.89\columnwidth]{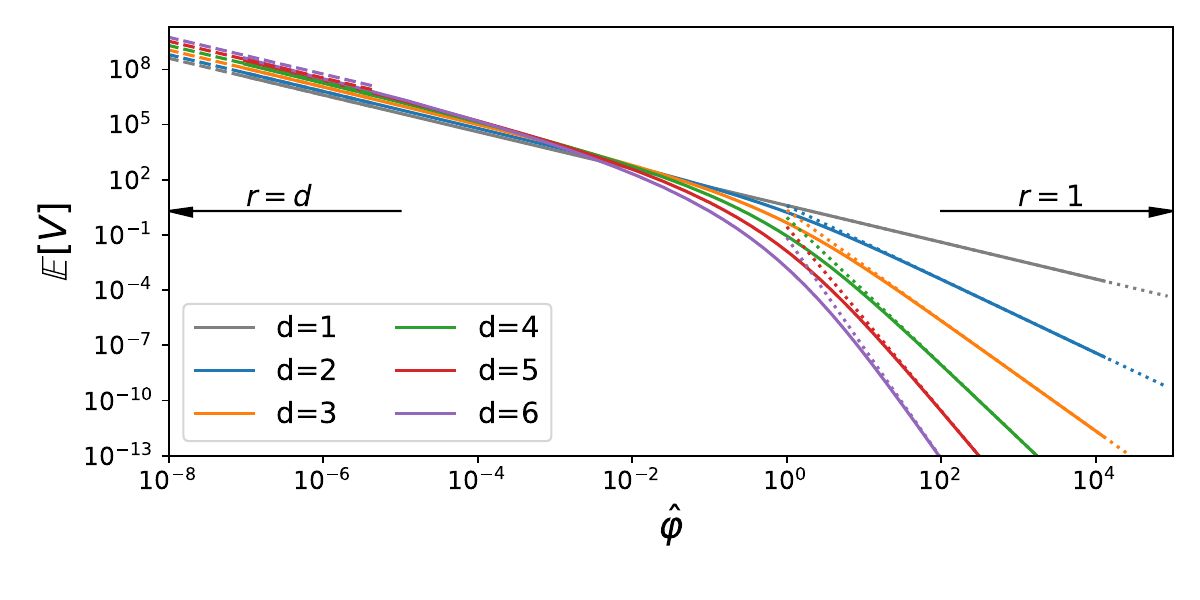}\\
		\caption{Average volume of the hRLG cell versus $\widehat{\varphi}$ for different $d$. The results agree with the exact PHT results for $r=d$ (dashed line) and $r=1$ (dotted line)~\cite{horrmann_volume_2014} at low and high densities, respectively.}  \label{fig:Volumes}
\end{figure}

\subsection{Delaunay triangulation: cell volume and its second-moment} 
Once the set of $N_v$ vertices has been obtained, a Delaunay triangulation--also implemented in the qhull package~\cite{barber1996quickhull}--is used to decompose the polytope into a set $\text{S}_k$ of $N_s$ polytopes with $d+1$ facets, i.e., simplexes. This decomposition, which is univocal and dual to the Voronoi tessellation,  can be algorithmically obtained in a time $N_v \log N_v$, which results in the complexity scaling at least exponentially in $d$. 

Given the set of simplexes $\text{S}_k$, the total volume of the cell is obtained by adding the volumes of each simplex (obtained through the determinant formula). Results are presented in Fig.~\ref{fig:Volumes}. The same is true for the moment of inertia distribution (or mean squared displacement) $\Delta=\sum^{N_s}_{k}[\mathbf{x}_{c,k}^2+\int_{V_{\text{S}_k}} (\mathbf{x}-\mathbf{x}_{c,k})^2/V_{\text{S}_k}]$, which can be simply evaluated around the center of mass $\mathbf{x}_{c,k}$ of each simplex~\cite{conway1982}.

\begin{figure}[t]
    \centering
		\includegraphics[width=0.49\columnwidth]{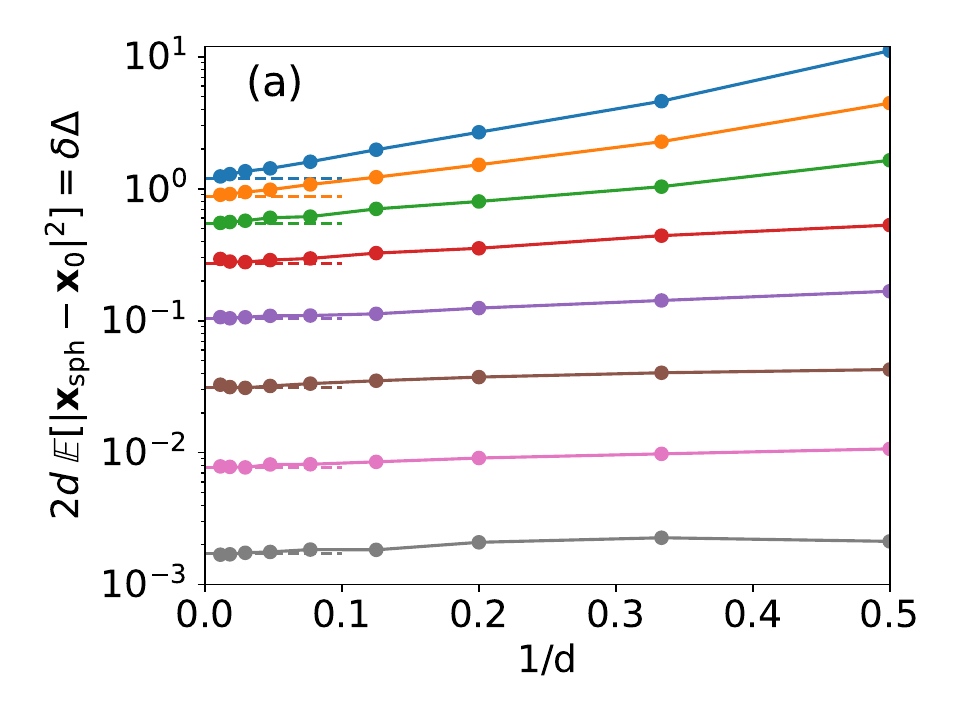}
        \includegraphics[width=0.49\columnwidth]{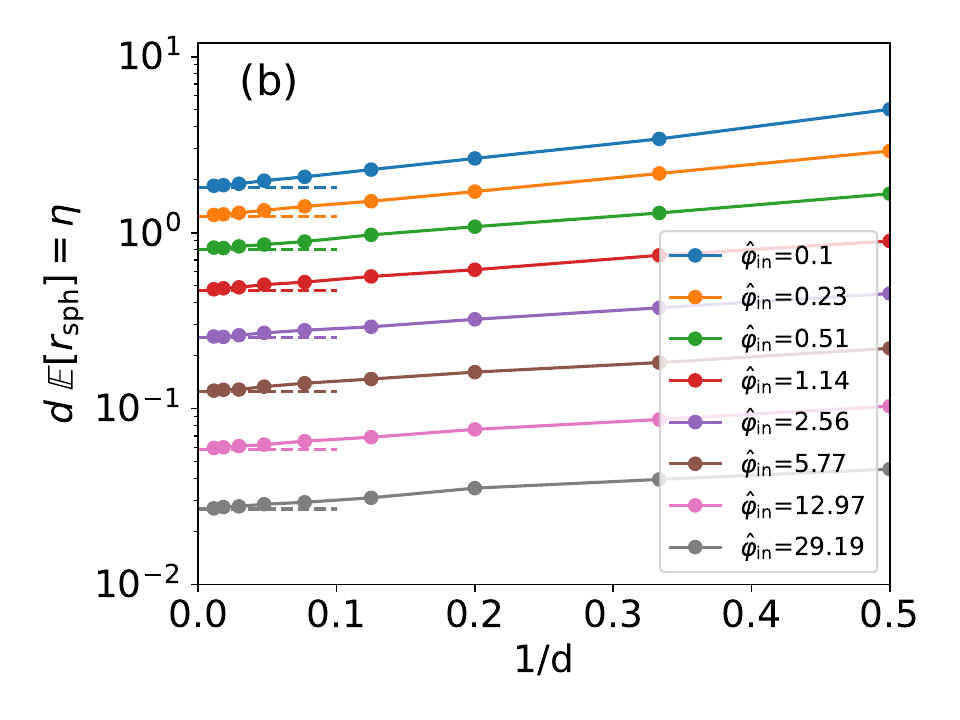}
		\caption{Dimensional scaling of the center $\mathbf{x}_\mathrm{sph}$ and radius $r_\mathrm{sph}$ of the inscribed sphere in the hRLG obtained from linear optimization. Initial packing fractions $\widehat{\varphi}_\mathrm{in}=0.1,\dots,29.19$ are considered for $d=2,\dots,89$ (following a Fibonacci sequence). Each result is averaged over 1000 realizations. \textbf{(a):} Dimensionally-rescaled mean squared distance of the center of the sphere from the origin, $\delta\Delta$. \textbf{(b):} Rescaled sphere radius $d r_\mathrm{sph}$. Dashed lines denote analytical results for the limit $d\to\infty$.}  \label{fig:dDeltaEta}
\end{figure}

\subsection{Convex optimization: insphere} 
Solving for the cell insphere can be formulated as a linear optimization problem
\begin{equation}\label{eq:LO}
    \max r,\quad \text{s.t. }\quad  \mathbf{V}_i  \cdot  \mathbf{x} + |\mathbf{V}_i|r\leq H_i \quad \text{for} \quad i = 1\dots M \ ,
\end{equation}
which finds the optimal sphere radius $r_\mathrm{sph}=r^*$ and its (Chebychev) center $\mathbf{x}_\mathrm{sph}$. Recall that the center of mass at the end of the compression algorithm described in the main text converges to $\mathbf{x}_\mathrm{sph}$.
The best linear optimization algorithm for Eq.~\eqref{eq:LO} scales linearly in $M$ (at fixed $d$) \cite{megiddo1984}. One might therefore expect complexity again to scale at least exponentially with $d$. However, because the (isostatic) insphere only touches $d+1$ hyperplanes, we no longer need to sample $M\gg\#\mathrm{facets}$ hyperplanes. A smaller $M$ (and correspondingly smaller $h_\mathrm{max}$) can instead be used. In practice, we find that $M=1000 d$ suffices to approach the $d\rightarrow\infty$ result. Within a few hours of CPU time, 1000 samples therefore can be obtained for up to $d=100$. Figure~\ref{fig:dDeltaEta} presents the results for the radius and center of the inscribed sphere obtained by solving the linear optimization problem. 

\begin{figure}[t]
    \centering
		\includegraphics[width=0.79\columnwidth]{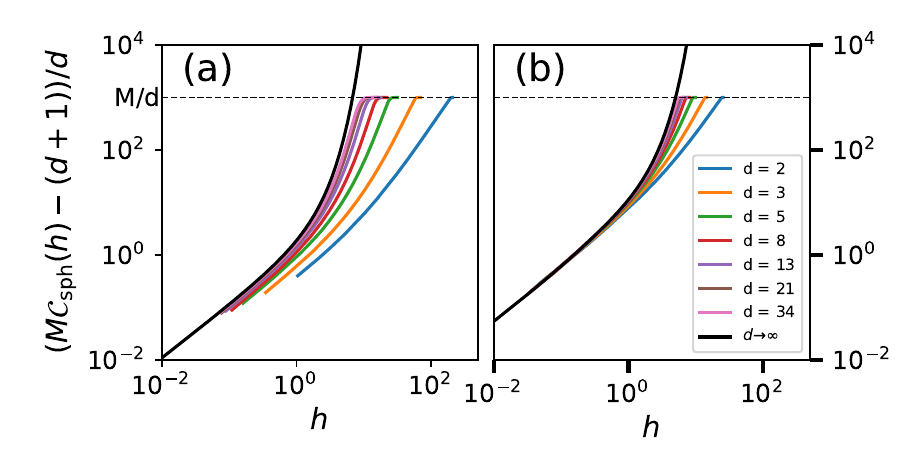}\\
		\caption{Cumulative distribution $\mathcal{C}_\mathrm{sph}$ of plane distances around the insphere. Simulation results are averaged over 1000 replicates for $d=2,\dots,34$ (following a Fibonacci sequence) at (a) $\hat\varphi=0.1$ and (b) $\hat\varphi=5.77$. The isostatic term is subtracted from the distribution to highlight the scaling near the sphere surface. The results converge to the analytical prediction of Eq.~\eqref{eq:gap_plus} in the limit $d\to\infty$ (solid black line). The distribution is cut off by fixing $M = 1000 d$ (dashed line).}  \label{fig:gaps}
\end{figure}

Given the center of the inscribed sphere, $\mathbf{x}_\mathrm{sph}$, it is also possible to evaluate the distribution of planes 
\begin{equation}
    P_\mathrm{sph}(h) = \mathbb{E}\big [\frac{1}{M}\sum_{i}^M\delta(h - h_{\mathrm{sph},i})\big ] \qquad \text{with}\quad h_{\mathrm{sph},i} = d(H_i-\mathbf{V}_i\cdot\mathbf{x}_\mathrm{sph}-r_\mathrm{sph}) \ . 
\end{equation}
If the inscribed sphere is isostatic (with only $d+1$-planes in contact), this distribution is expected to take the form $M P_\mathrm{sph}(h) = (d+1)\delta(h)+\rho^{+}(h)/d$, where $\rho^{+}(h)$ is a smooth function. Simulation results in Fig.~\ref{fig:gaps} show that $\rho^{+}(h)= \text{const}$ at small $h$, consistently with the analytical predictions for the limit $d\to\infty$ obtained in Sec.~\ref{app:gap}. Note that $\rho^{+}(h)$ grows linearly at small $h$, thus resulting in a  critical structural exponent that is trivial.

\section{Geometrical crossover}\label{app:cross}
\begin{figure}[t]
    \centering
		\includegraphics[width=0.89\columnwidth]{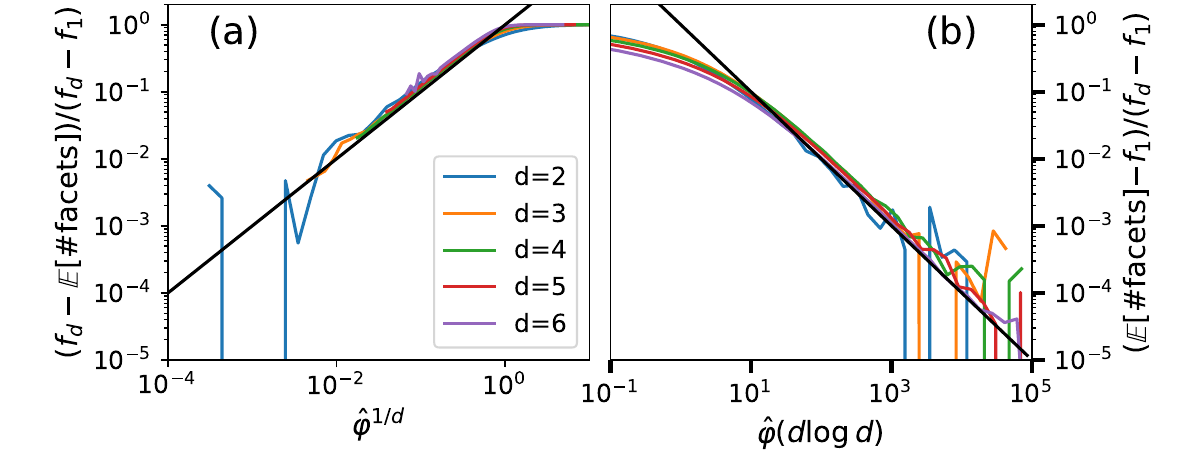}\\
		\caption{Convergence of the number of facets to (a) the super-exponential and (b) the exponential regimes from simulations with $d=2\dots6$, where $f_d$ is the exact number of facets in the PHT model with $r=d$. The black lines denote $x$ and $x^{-1}$, respectively. Note that the ($y$-axis) distance from the exact values is renormalized with $f_d-f_1$.}  \label{fig:Cross}
\end{figure}

As shown in Fig.~\ref{fig:Limits}, both the average number of facets $\mathbb{E}[\#\mathrm{facets}]$ and the average number of vertices $\mathbb{E}[\#\mathrm{vertices}]$ exhibit a crossover from an exponential $\exp(d)$ to a super-exponential $d\exp(d)$ scaling as density is reduced. Two lines of numerical evidence suggest that the crossover converges to a sharp transition at $\widehat{\varphi}=0$ in the limit of $d\to\infty$. 
First, because the convergence to the super-exponential regime scales as $\widehat{\varphi}^{1/d}$ (Fig.~\ref{fig:Cross}a), to have half the facets expected at $\widehat{\varphi}=0$ one 
needs $\widehat{\varphi}=1/2^d$, which fast converges to $0$. Second, because the approach to the exponential regime  goes as $\widehat{\varphi}^{-1}(d\log d)$ (Fig.~\ref{fig:Cross}b), to have half of the facets expected at $\widehat{\varphi}=0$ one needs $\widehat{\varphi}=\widehat{\varphi}=1/2/(d\log d)$, which also converges (albeit more slowly) to $0$ with increasing $d$. 

\section{\lowercase{h}RLG replica formulation}

In order to obtain analytical results for the limit $d\rightarrow\infty$, we start from the replicated partition function
\begin{equation}\label{eq:Free}
\mathbb{E}[Z^n] = \exp\{-\lambda \int d\mathbf{R} [e^{-\beta V(\mathbf{R})} -1]\} \int d^n\mathbf{x} \exp \Big \{\lambda \int d\mathbf{R} [e^{-\beta V(\mathbf{R})}e^{-\beta \sum_{a=1}^n W(-\mathbf{x}_{a};\mathbf{R})} - 1]\Big \} \ ,
\end{equation}
where $\beta V(\mathbf{R})$ is the radial function that describes the distribution of obstacles and $W(-\mathbf{x};\mathbf{R})$ is a function that depends on the projection of $\mathbf{x}$ in the direction defined by $\mathbf{R}$. In particular, it is a function of the distance from the ``tangent'' plane 
\begin{equation}
    \frac{\mathbf{R}}{|\mathbf{R}|} \cdot \mathbf{x} = |\mathbf{R}|-l
\end{equation}
with $l$ the radius of the obstacle centered at $\mathbf{R}$. This plane is tangent to the spherical obstacle (centered at $\mathbf{R}$ and of radius $l$) and orthogonal to $\mathbf{R}$ and is equivalent to Eq.~\eqref{eq:half} with $H=|\mathbf{R}|-l$ and $\mathbf{V}=\frac{\mathbf{R}}{|\mathbf{R}|}$.
For hard spheres (HS) and hard planes (HP), we have respectively
\begin{align}
    e^{-\beta V(\mathbf{R})} &= \theta(|\mathbf{R}|>l)\\
    e^{-\beta W(-\mathbf{x};\mathbf{R})} &= \theta(\frac{\mathbf{R}}{|\mathbf{R}|} \cdot \mathbf{x} < |\mathbf{R}|-l) \ .
\end{align}
In the limit $d\to\infty$, because $\lambda=d \widehat{\varphi}/V_d$, the measures in Eq.~\eqref{eq:Free} concentrate on the saddle point in $\mathbf{R}$. It is then possible to study the replicated Mayer function
\begin{equation}
\label{eq:mayer}
    f^n(\mathbf{R}) = \int d^n\mathbf{x} \Big \{ e^{-\beta V(\mathbf{R})}e^{-\beta \sum_{a=1}^n W(-\mathbf{x}_{a};\mathbf{R})} - 1\Big \},
\end{equation}
which is by definition a large deviation function of $\lambda$.
In order to evaluate Eq.~\eqref{eq:mayer}, we assume that in the limit $d\to\infty$ the distribution of $\mathbf{x}_a$ is Gaussian and concentrates in a shell $1/d$ around the typical value, in analogy with~\cite[Eq.~(4.44)]{book_2020}.
This scaling regime is examined by considering the rescaled radius $h = d H$, given by $|\mathbf{R}|=l(1+h/d)$, thus defining the shell potentials
\begin{align}
    \bar{v}(h) &\equiv V \big ( l(1+\frac{h}{d}) \big )  \\
    \bar{w}(h) &\equiv W \big (l \frac{h}{d}; R \big ) 
\end{align}
We then assume that $\mathbf{x}_a$ is a rotationally invariant (in $\mathbb{R}^d$) Gaussian distribution with zero mean $\mathbb{E}[\mathbf{x}_a]=0$ and variance
\begin{equation}\label{eq:var}
    \mathbb{E}[\mathbf{x}_a \cdot \mathbf{x}_b] = \frac{l^2}{d}\alpha_{ab} \qquad \forall \; a,b
\end{equation}
where the $l^2$ prefactor is given for convenience and $d^2$ is the presumed scaling.
To consider $W$ in this regime, we define the distance from the hyperplane
\begin{equation}
    y_a = \frac{\mathbf{R}}{|\mathbf{R}|} \cdot \mathbf{x}_a - (|\mathbf{R}|-l),
\end{equation}
with mean $\mathbb{E}[y_a] = - (|\mathbf{R}|-l)$ and variance given by Eq.~\eqref{eq:var}. Rescaling then gives
\begin{equation}
    \frac{d}{l}y_a = -h + z_a \qquad \forall \; a ,
\end{equation}
where $h$ is the rescaled radius and $z_a$ is a normal distribution with variance $\alpha_{ab}$.
We finally obtain
\begin{equation}\begin{split}
\overline{f^n}(h) &= e^{-\beta \bar{v}(h)}e^{\sum_{a,b=1}^{n}\frac{\alpha_{ab}}{2}\partial_{h_a}\partial_{h_b}} e^{-\beta\sum_{c=1}^n\bar{w}(h_c)}\big|_{h_c=h}-1 \ .
\end{split}\end{equation}
The only difference from the RLG result is that the diagonal term $\alpha_{aa}$ is missing. Because this term encodes for the curvature of the obstacles  (see \cite[Eq.~(4.52)]{book_2020} and \cite[Eq.~(133)]{Fol2022}), it is properly expected to vanish for hyperplanes. The resulting free energy is then
\begin{equation}\label{eq:rep_free}
    \log(\mathbb{E}[Z^n]) = n\frac{d}{2}\log(\frac{\pi e}{d^2}) +\frac{d}{2}\log\det\mathbb{\alpha} + d\hat\varphi \int_{-\infty}^{\infty} dh e^h \overline{f^n}(h) \ .
\end{equation}

\subsection{RS solution}
We then rewrite the overlap matrix $\alpha_{ab}$ in terms of mean squared displacements (MSD)
 \begin{equation}
     \Delta_{ab} = \frac{d}{l^2}\mathbb{E}[|\mathbf{x}_a - \mathbf{x}_b|^2] = \alpha_{aa}-2\alpha_{ab}+\alpha_{bb} \ .
 \end{equation}
In the following we set $l=1$ for simplicity.
Assuming that the MSD has the \emph{planted} replica symmetric (RS) structure
\begin{equation}
\Delta_{ab} = 
\begin{cases}
    0 \quad &\text{for} \quad a=b\\
    \Delta_r \quad &\text{for} \quad  a=0,b\neq0 \wedge a\neq0,b=0 \\
    \Delta \quad &\text{for} \quad a\neq0 \vee b\neq0 \vee a\neq b\\
\end{cases}
\qquad
\text{e.g., for $n=3$}\quad
     \begin{pmatrix}
         0 & \Delta_r & \Delta_r & \Delta_r \\
         \Delta_r & 0 & \Delta & \Delta \\
         \Delta_r & \Delta & 0 & \Delta \\
         \Delta_r & \Delta & \Delta & 0 \\
     \end{pmatrix}.
 \end{equation}
where $\Delta_r$ is the MSD with the origin while $\Delta$ is the MSD between a pair of replicas (see \cite[Eq.~(4.63)]{book_2020}). Given this $(n+1)\times (n+1)$ matrix of MSD, the $n\times n$ matrix of overlaps reads
\begin{equation}
    \alpha_{ab}=\frac{1}{2}\big ( \Delta \delta_{ab} + \delta\Delta \big ) \ ,
\end{equation}
where we introduce the new variable $\delta\Delta = 2\Delta_r-\Delta$.
Note that the quantities $\Delta$ and $\delta\Delta$ have the following geometrical meaning
\begin{equation}
    \Delta=2d \; \frac{\int_{V_\mathrm{cell}} d\mathbf{x} (\mathbf{x}-\mathbf{x}_\mathrm{c})^2}{\int_{V_\mathrm{cell}} d\mathbf{x}} \qquad \delta\Delta = 2d \; \mathbf{x}_\mathrm{c}^2
\end{equation}
with $V_\mathrm{cell}$ the volume of the cell and $x_\mathrm{c}$ its center of mass.

\begin{figure}[t]
    \centering
		\includegraphics[width=0.79\columnwidth]{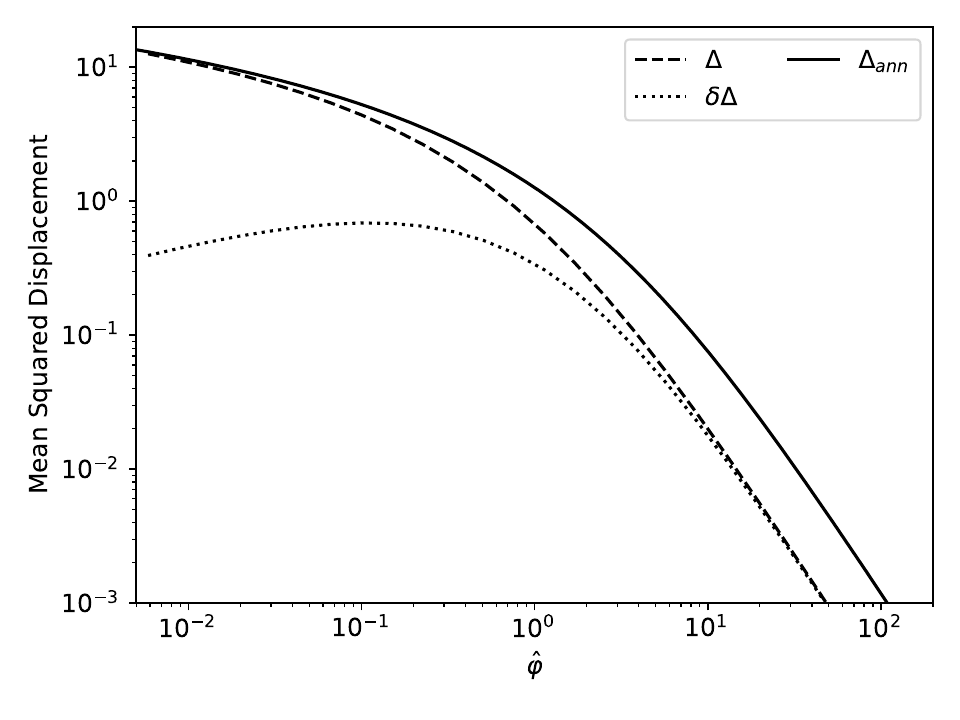}\\
		\caption{Quenched $\Delta,\delta\Delta$ and annealed $\Delta_\mathrm{ann}$, as function of the packing fraction $\widehat{\varphi}$. }  \label{fig:Deltas}
\end{figure}

Given this RS structure, the quenched free energy $\mathbb{E}[\log(Z)]=\lim_{n\to 0} \partial_n \log(\mathbb{E}[Z^n])$ can be evaluated. The second term of Eq.~\eqref{eq:rep_free} is
\begin{equation}
    \frac{d}{2}\lim_{n\to0}\partial_n\log\det\mathbb{\alpha} = \frac{d}{2} \big ( \log(\Delta)+\frac{\delta\Delta}{\Delta} \big )
\end{equation}
and the third term is 
\begin{equation}\label{freeRS}
\begin{aligned}
    d\widehat{\varphi} \lim_{n\to0}\partial_n  \int_{-\infty}^{\infty} dh e^h \overline{f^n}(h) &= d\widehat{\varphi} \lim_{n\to0}\partial_n \int_{-\infty}^{\infty} dh e^h \big [  e^{-\beta \bar{v}(h)} e^{\frac{\delta \Delta}{4}\partial^2_h} (e^{\frac{\Delta}{4}\partial^2_h}e^{-\beta \bar{w}(h)})^n - 1 \big ]\\
    & = d\widehat{\varphi} \int_{-\infty}^{\infty} dh e^h   e^{-\beta \bar{v}(h)} e^{\frac{\delta \Delta}{4}\partial^2_h} \log(e^{\frac{\Delta}{4}\partial^2_h}e^{-\beta \bar{w}(h)})\\
    & = d\widehat{\varphi} \int_{-\infty}^{\infty} dh e^h   e^{-\beta \bar{v}(h)} e^{\frac{\delta \Delta}{4}\partial^2_h} f(h)
\end{aligned}
\end{equation}
where $f(h)=\log(e^{\frac{\Delta}{4}\partial^2_h}e^{-\beta \bar{w}(h)})$. We then introduce the function
\begin{equation}
g_\mathrm{RLG}(h)=e^{\frac{\delta\Delta}{4}\partial^2_h}e^{-\beta \bar{v}(h+\frac{\delta\Delta}{4})},
\end{equation}
which describes the correlations with the origin (the result is the same for the RLG, hence the subscript),
and
\begin{equation}\label{eq:gpl}
g_\mathrm{PL}(h)=e^{\frac{\Delta}{4}\partial^2_h}e^{-\beta \bar{w}(h-\frac{\delta\Delta}{4})} \ .
\end{equation}
which describes correlations with the hyperplanes (PL). The fact that these two function are different reflects the asymmetry of the problem. (In the RLG, the two functions are the same because of the symmetry between the tracer and the spherical obstacles.) Using $g_\mathrm{RLG}(h),g_\mathrm{PL}(h)$ (by means of simple manipulations, see \cite[Sec.~4.3]{book_2020}) the free energy reads 
\begin{equation}\label{eq:freeF}
\mathbb{E}[\log(Z)] = \frac{d}{2}\log(\frac{\pi e}{d^2}) +\frac{d}{2} \big ( \log(\Delta)+\frac{\delta \Delta}{\Delta} \big ) + d \widehat{\varphi} \int_{-\infty}^{\infty} dh e^h g_\mathrm{RLG}(h) \log \big( g_\mathrm{PL} (h) \big)
\end{equation}
After extremizing with respect to $\Delta$ and $\delta\Delta$, two coupled equations are obtained
\begin{eqnarray}\label{eq:sol}
    \delta\Delta = 2 \Delta^2 \widehat{\varphi} \int_{-\infty}^{\infty} dh e^h g_\mathrm{RLG}(h) \frac{1}{4} f'(h-\delta\Delta/4)^2 = 2 \Delta^2 \widehat{\varphi} \int_{-\infty}^{\infty} dh e^h g_\mathrm{RLG}(h) \frac{1}{4} \Big ( \frac{\partial_h g_\mathrm{PL} (h)}{ g_\mathrm{PL} (h)} \Big )^2 \\\label{eq:sol2}
    \Delta^{-1} = -2\widehat{\varphi} \int_{-\infty}^{\infty} dh e^h g_\mathrm{RLG}(h) \frac{1}{4} f''(h-\delta\Delta/4) = -2\widehat{\varphi} \int_{-\infty}^{\infty} dh e^h g_\mathrm{RLG}(h) \frac{1}{4} \Big [ \frac{\partial^2_h g_{PL} (h)}{ g_\mathrm{PL} (h)} -\big ( \frac{\partial_h g_\mathrm{PL} (h)}{ g_\mathrm{PL} (h)}\big )^2  \Big ] \ .
\end{eqnarray}
Focusing on the hard-sphere/hard-plane case (HS/HP), we have
\begin{eqnarray}
    g_\mathrm{RLG,HS}(h)=\frac{1}{2}\big ( 1 + \text{erf}(\frac{h+\frac{\delta \Delta}{4}}{\sqrt{\delta \Delta}}) \big )\\
    g_\mathrm{PL,HP}(h)=\frac{1}{2}\big ( 1 + \text{erf}(\frac{h-\frac{\delta \Delta}{4}}{\sqrt{\Delta}}) \big ) \label{eq:PL}
\end{eqnarray}
which has to be plugged into Eqs.~\eqref{eq:sol} and \eqref{eq:sol2}. The resulting equations can be recursively solved by numerical integration, thus providing $\Delta(\widehat{\varphi})$ and $\delta\Delta(\widehat{\varphi})$ as a function of the packing fraction $\widehat{\varphi}$. 
The resulting RS solution is stable over the whole range of $\widehat{\varphi}\in[0,\infty]$ (the replicon eigenvalue being always positive). As shown in Fig.~\ref{fig:Deltas}, while $\Delta$ diverges as $\widehat{\varphi} \to 0$, $\delta\Delta(\varphi)\to 0$.
Note that because $\delta\Delta\neq 0$ the origin of the cell (with MSD $\Delta_r$) and the center of the cell (with MSD $\Delta$) do not coincide. As shown in Sec.~\ref{sec:annealed}, $\delta\Delta(\varphi)=0$ corresponds to the annealed solution. Note that these equations can be studied for different potentials $V(\mathbf{R})$ and $W(-\mathbf{x};\mathbf{R})$, and therefore for different $P(H)$, as long as the potentials concentrate to some definite functions $\bar{v}(h)$ and $\bar{w}(h)$ in the limit $d\to\infty$~\cite[Sec.~2.3.2]{book_2020}.

\subsection{Annealed solution}
\label{sec:annealed}
Reproducing the same calculations for the unreplicated system (setting $n=1$ from the start), we obtain the annealed approximation
\begin{equation}
\log(\mathbb{E}[Z]) =  \frac{d}{2}\log(\frac{\pi e}{d^2}) +\frac{d}{2} \log(\Delta) + d \widehat{\varphi} \int_{-\infty}^{\infty} dh e^h \big( e^{-\beta \bar{v}(h)}  g_\mathrm{PL} (h) - 1\big)
\end{equation}
with $\delta\Delta = 0$. After extremizing with respect to $\Delta$, the annealed solution is obtained
\begin{equation}
    \widehat{\varphi}^{-1} = -2 \Delta \int_{-\infty}^{\infty} dh e^h e^{-\beta \bar{v}(h)} \frac{1}{4} \partial^2_h g_\mathrm{PL} (h) \ .
\end{equation}
If we consider the hard plane case from Eq.~\eqref{eq:PL}, then
\begin{equation}
    \widehat{\varphi}^{-1} = \Delta \int_{0}^{\infty} dh e^h \frac{h e^{-\frac{h^2}{\Delta }}}{ \sqrt{\pi } \Delta ^{3/2}} = \frac{\sqrt{\Delta }}{2 \sqrt{\pi }}+\frac{1}{4} e^{\Delta /4} \Delta  \left( 1 + \text{erf}\left(\frac{\sqrt{\Delta }}{2}\right)\right)\ ,
\end{equation}
which implicitly defines $\Delta$ as a function of $\widehat{\varphi}$. Recalling that for hard planes the partition function is equivalent to the volume of the cell, the logarithm of the annealed volume of a cell can then be evaluated as a function of its annealed $\Delta$ 
\begin{equation}
\frac{1}{d}\log(\mathbb{E}[V_\mathrm{cell}])+\log(d) =  \frac{1}{2}\log(\pi e \Delta) +  \frac{\widehat{\varphi}}{2} \left(1-e^{\Delta /4}\left(1+\text{erf}\left(\frac{\sqrt{\Delta }}{2}\right)\right) \right)
\end{equation}
Figure~\ref{fig:Deltas} shows $\Delta$ as a function $\widehat{\varphi}$,  compared with the quenched result, and  Fig.~\ref{fig:AnnVol} shows the volume versus $\widehat{\varphi}$,  compared with the $d\rightarrow\infty$ extrapolation of finite $d$ simulation results. \\

\subsection{State following and jamming}
The RS free energy can be continuously tracked upon changing the cell shape while keeping its origin fixed through state following. Here, we specifically wish to compress the cell so that every hyperplane is $\eta/d$-shifted with respect to the origin. This can be achieved by simply substituting $\bar{w}(h) \to \bar{w}(h-\eta)$. Increasing $\eta$ reduces the volume of the cell, and for every initial density $\widehat{\varphi}_\mathrm{in}$ there exists a maximum $\eta_\mathrm{J}$ such that the cell reduces to a single point, which corresponds to the jamming limit. The limiting cell is expected to be a simplex with $d+1$ facets and therefore to be isostatic. Put differently, the final point $\mathbf{x}_\mathrm{sph}$ is the center of the largest inscribed sphere, with radius $r_\mathrm{sph}$, in the original cell.

The RS equations for the state followed (compressed) cell are the same as Eq.~\eqref{eq:sol} after substituting $g_\mathrm{PL}(h)$ from Eq.~\eqref{eq:gpl} by $g_\mathrm{PL}(h-\eta)$ with the parameter $\eta$ controlling the compression.
The resulting equations can then be numerically solved to obtain $\Delta$ and $\delta\Delta$ for given $\widehat{\varphi}_\mathrm{in}$ and $\eta$. The compression protocol consists of increasing $\eta$ until at $\eta_\mathrm{j} =d \mathbb{E}[r_\mathrm{sph}]$ the cell contracts to a single point, and therefore $\Delta_\mathrm{j} = 0$, while $\delta\Delta_\mathrm{j}=2 d \mathbb{E}[|\mathbf{x}_\mathrm{sph}-\mathbf{x}_\mathrm{0}|^2]$ remains finite.
Figure~\ref{fig:SF} shows the compression lines for different starting $\widehat{\varphi}_\mathrm{in}$.
The end points $\eta_\mathrm{j}(\widehat{\varphi}),\delta\Delta_\mathrm{j}(\widehat{\varphi})$ of different compression protocols for $\widehat{\varphi}\in[0,\infty]$ define the jamming line, which in the hRLG can be reached without encountering any instability, such as a Gardner transition. In other words, the RS solution is stable up to the jamming line for any $\widehat{\varphi}$. The stability is a direct consequence of the convexity of the initial polytope.

\subsection{Jamming line with harmonic planes}
In order to characterize the states on the jamming line, we adopt the overjammed calculation scheme. In this scheme, HP are replaced by harmonic planes (hrmP)
\begin{equation}
\bar{w}(h) = \frac{h^2}{2}\theta(-h) \ .
\end{equation}
By definition the limit $T\to0$ for hrmP is equivalent to the HP case whenever the HP cell exists, but the jamming transition can then be studied starting from the overjammed phase ($\eta\to\eta^+_\mathrm{j}$), i.e., when hrmP superimpose (UNSAT phase). The advantage of the harmonic model is that in this limit, the number of contacts corresponds to the number of half-spaces that intersect at $\mathbf{x}_\mathrm{sph}$, which is countable even before jamming is reached. 

\begin{figure}[t]
    \centering
		\includegraphics[width=0.79\columnwidth]{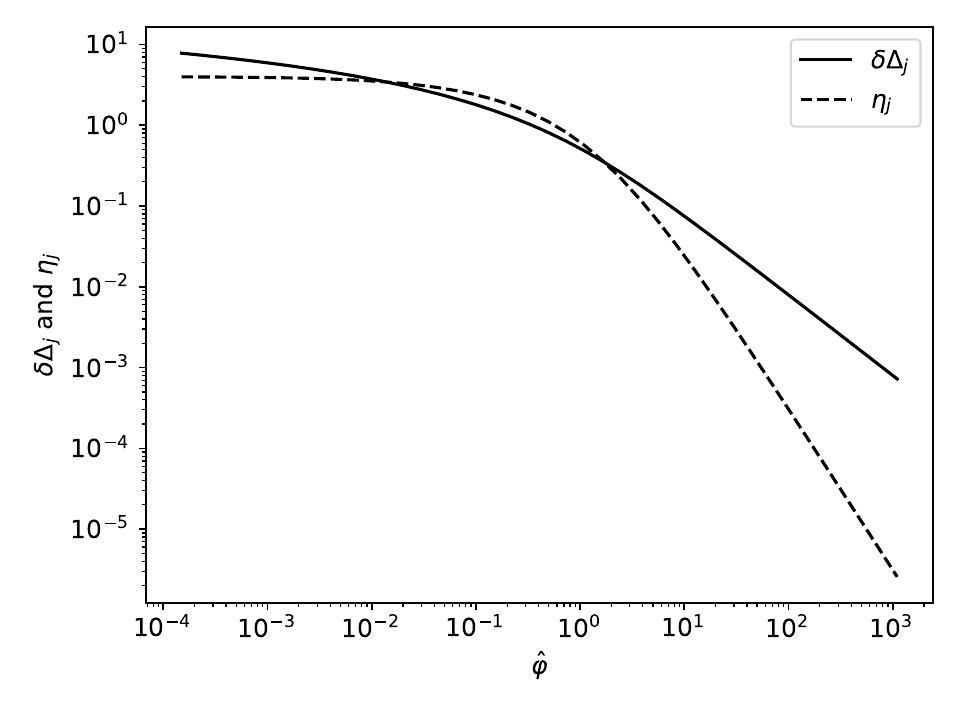}
		\caption{Values of $\eta_\mathrm{j}$ and $\delta\Delta_\mathrm{j}$ along the jamming line. For $\widehat{\varphi}\rightarrow\infty$, $\eta_\mathrm{j}$ converges to 4.} \label{fig:JLine}
\end{figure}

The free energy of the model is given in Eq.~\eqref{eq:freeF}. To study the limit $T\to0$, we consider the change of variable $\frac{\Delta}{2} = \chi T$, which is expected to be a reasonable scaling in the overjammed phase \cite[Sec.~9.4.3]{book_2020}. Using a saddle point approximation with parameter $\beta=1/T$, we can then rewrite (see \cite[Appendix C]{franz2017})
\begin{equation}
f_0(h)\equiv\lim_{T\to0,\Delta=\chi T} \log(g_\mathrm{PL}(h)) = -\frac{1}{T(1+\chi)}\frac{h^2}{2}\theta(-h),
\end{equation}
where for state following $h\to h-\eta$. The free energy in the limit $T\to0$ transforms into 
\begin{equation}
\mathbb{E}[\log(Z)] = \text{const} +\frac{d}{2} \big ( \log(\chi T)+\frac{\delta \Delta}{2\chi T} \big ) - d \widehat{\varphi} \frac{1}{T(1+\chi)}\frac{1}{2}\int_{-\infty}^{\eta+\frac{\delta\Delta}{4}} dh e^h g_\mathrm{RLG}(h)  \big( h-\eta-\frac{\delta\Delta}{4} \big)^2 \ .
\end{equation}
Extremizing with respect to $\chi$ and $\delta\Delta$ gives  two coupled equations
\begin{eqnarray}
    \delta\Delta = 2\widehat{\varphi}\big(\frac{\chi}{1+\chi}\big)^2\int_{-\infty}^{\eta+\frac{\delta\Delta}{4}} dh e^h g_\mathrm{RLG}(h) \big( h-\eta-\frac{\delta\Delta}{4} \big)^2 = 2\chi^2e_\mathrm{RS}\\
    \label{eq:dD}
    1 = \widehat{\varphi}\frac{\chi}{1+\chi}\int_{-\infty}^{\eta+\frac{\delta\Delta}{4}} dh e^h g_\mathrm{RLG}(h) \label{eq:1}
\end{eqnarray}
where $e_\mathrm{RS}=-\partial_\beta\mathbb{E}[\log(Z)]/d$ is the energy. Recall that $g_\mathrm{RLG}(h)$ contains $\delta\Delta$. Approaching the jamming transition from below, given Eq.~\eqref{eq:dD} (and given that $\delta\Delta$ is finite and $e_{RS}\rightarrow0$) the susceptibility must diverge ($\chi \to \infty$) and the resulting equations for the jamming line are,
\begin{eqnarray}
    \delta\Delta_\mathrm{j} = 2\widehat{\varphi}\int_{-\infty}^{\eta_\mathrm{j}+\frac{\delta\Delta_\mathrm{j}}{4}} dh e^h \frac{1}{2}\Big(1+\text{erf}\big (\frac{4h-\delta\Delta_\mathrm{j}}{\delta\Delta_\mathrm{j}}\big ) \Big ) \big( h-\eta_\mathrm{j}-\frac{\delta\Delta_\mathrm{j}}{4} \big)^2\\
    1 = \widehat{\varphi}\int_{-\infty}^{\eta_\mathrm{j}+\frac{\delta\Delta_\mathrm{j}}{4}} dh e^h \frac{1}{2}\Big(1+\text{erf}\big (\frac{4h-\delta\Delta_\mathrm{j}}{\delta\Delta_\mathrm{j}}\big ) \Big ).
\end{eqnarray}
This expression implicitly defines $\eta_\mathrm{j}(\widehat{\varphi})$ and $\delta\Delta_j(\widehat{\varphi})$, which are respectively equal to $d \mathbb{E}[r_\mathrm{sph}]$ and $2d \mathbb{E}[|\mathbf{x}_\mathrm{sph}-\mathbf{x}_\mathrm{0}|^2]$. The results are shown in Fig.~\ref{fig:JLine} and are compared with simulations in Fig.~\ref{fig:dDeltaEta}.

\subsection{Gap probability distribution and isostaticity}\label{app:gap}

The gap probability distribution $\rho(h)$ of the planes in the state-followed cell can be evaluated by functional differentiation of the free energy \ref{freeRS},
\begin{equation}
    \begin{aligned}
        \rho(h) &= \frac{\delta  \mathbb{E}[\log(Z)] }{\delta (-\beta w(h))} 
        = d\widehat{\varphi} e^h\bar{g}(h) \\
        &= d\widehat{\varphi} \int_{-\infty}^{\infty} dz e^{z+\eta}   e^{-\beta \bar{v}(z+\eta)} e^{\frac{\delta \Delta}{4}\partial_{z^2}} \big [(e^{\frac{\Delta}{4}\partial_{z^2}}e^{-\beta \bar{w}(z)})^{-1} e^{\frac{\Delta}{4}\partial_{z^2}}\frac{\delta e^{-\beta \bar{w}(z)}}{\delta (-\beta w(h))}\big ]\\
    \end{aligned}
\end{equation}
where $\bar{g}(h)$ is the two-point correlation function.
Integrating the last line by parts, we obtain the RS correlation function (equivalent to \cite[Eq.~(6.19)]{book_2020})
\begin{equation}
e^h\bar{g}(h) = e^{-\beta \bar{w}(h)} e^{\frac{\Delta}{4}\partial^2_h}  \big [ \frac{e^{\frac{\delta\Delta}{4}\partial^2_h}
e^{h+\eta - \beta \bar{v}(h+\eta)}  }{e^{\frac{\Delta}{4}\partial^2_h}e^{-\beta \bar{w}(h)}} \big ]
\end{equation}
Considering HP, the small cell limit of the correlation function reads~\cite[Eq.~(9.71)]{book_2020})
\begin{equation}
\begin{aligned}\label{eq:gap_plus}
    \rho^{+}(h)/d \equiv \hat\varphi\lim_{\Delta\to 0} e^h\bar{g}(h) &= \hat\varphi\;\theta(h)e^{\frac{\delta\Delta}{4}\partial^2_h}
e^{h+\eta - \beta \bar{v}(h+\eta)}\\
&= \hat\varphi\; \theta(h)e^{h+\eta+\frac{\delta\Delta}{4}}e^{\frac{\delta\Delta}{4}\partial^2_h}e^{- \beta \bar{v}(h+\eta+\frac{\delta\Delta}{2})}\\
&= \hat\varphi\; \theta(h)e^{h+\eta+\frac{\delta\Delta}{4}}g_{rlg}(h+\eta+\frac{\delta\Delta}{4})
\end{aligned}
\end{equation}
If we instead consider hrmP in the limit of $T\to0$, the positive gaps ($h>0$) have the same distribution of HP. This result is consistent with simulations (see Fig.~\ref{fig:gaps} on the jamming line with $\Delta_\mathrm{j}$ and $\eta_\mathrm{j}$ derived in the previous section). While for the negative gaps ($h<0$) we obtain 
\begin{equation}
\begin{aligned}
    \rho^{-}(h)/d \equiv \hat\varphi\lim_{T \to 0} e^h\bar{g}(h) &\approx \hat\varphi\;e^{-\beta \bar{w}(h)} e^{\frac{\chi T}{2}\partial^2_h}  \big [ \frac{e^{\frac{\delta\Delta}{4}\partial^2_h}
e^{h+\eta - \beta \bar{v}(h+\eta)}  }{e^{\frac{\chi T}{2}\partial^2_h}e^{-\beta \bar{w}(h)}} \big ]\\
&=  \hat\varphi\; \theta(-h)\int \frac{d k}{\sqrt{2\pi\chi T}} e^{-\frac{\beta}{2\chi} (h-k)^2}e^{\frac{\beta}{2(1+\chi)} k^2}\big [ e^{\frac{\delta\Delta}{4}\partial^2_k}
e^{k+\eta - \beta \bar{v}(k+\eta)}  \big ]\\
&= \hat\varphi\; \theta(-h)e^{h(1+\chi)+\eta+\frac{\delta\Delta}{4}}g_{rlg}(h(1+\chi)+\eta+\frac{\delta\Delta}{4})(1+\chi) \ ,
\end{aligned}
\end{equation}
where from the second to the third line we have used the saddle point in $k$ given the large parameter $\beta$ (see also \cite{franz2017,ikeda2020}). The total number of interacting planes (superimposed in $\mathbf{x}_\mathrm{sph}$) can then be found by integrating the two-point correlation for negative gaps,
\begin{equation}
z = \int_{-\infty}^{0}dh \rho(h) =d\widehat{\varphi}\int_{-\infty}^{\eta+\frac{\delta\Delta}{4}}dh e^h g_{rlg}(h)\\
= d\frac{1+\chi}{\chi} \ ,
\end{equation}
where the last equality uses Eq.~\eqref{eq:1}. On the jamming line $\chi\to\infty$ and the system is isostatic, i.e. $z=d$.
Note that this result differs from the usual hard-sphere isostaticity, because it does not require the presence of a prior Gardner phase from the HP side.

\subsection{High density limit: RLG $\to$ PHT($r=1$) }
Following \cite[Eq.~(7.36)]{book_2020}, we have that the typical logarithm of the volume of a cage (which for a hard sphere potential corresponds to the free energy) in the limit $d\to\infty$ is 
\begin{equation}\label{eq:vol_cage}
\mathbb{E}[\log(V_\mathrm{cage})]=
d \left (-\log(d) + \frac{1}{2}+\frac{1}{2}\log(2 \pi \Delta) + \widehat{\varphi} \int_{-\infty}^{\infty} dh e^h q_\mathrm{RLG}(\Delta;h)\log(q_\mathrm{RLG}(\Delta;h)) \right )  \ ,
\end{equation}
with 
\begin{equation}
q_\mathrm{RLG}(\Delta;h) = \frac{1}{2} \left ( 1 + \text{erf}\left(\frac{\frac{\Delta }{4}+h}{\sqrt{\Delta }}\right) \right ),
\end{equation}
where $\text{erf}(x) = \frac{2}{\pi}\int_0^{\infty}dx e^{-x^2}$ and $\Delta$ is the rescaled MSD, which is an implicit function of  $\widehat{\varphi}$,
\begin{equation}\label{eq:phi}
    \widehat{\varphi}^{-1}  = -2 \Delta \int_{-\infty}^{\infty} dh e^h \log ( q_\mathrm{RLG}(\Delta;h)))\partial_{\Delta} q_\mathrm{RLG}(\Delta;h)  = \int_{-\infty}^{\infty} dh \frac{e^{-\frac{\Delta }{8}-\frac{2 h^2}{\Delta }}}{\pi  \left(\text{erf}\left(\frac{\frac{\Delta }{4}+h}{\sqrt{\Delta }}\right)+1\right)}  \ .
\end{equation}

An analogous formula can be written for the logarithm of the typical volume, i.e., the annealed volume computation, 
\begin{equation}\label{eq:vol_cage_ann}
\log(\mathbb{E}[V_\mathrm{cage}])=
d \left (-\log(d) + \frac{1}{2}+\frac{1}{2}\log(2 \pi \Delta) + \widehat{\varphi} \int_{-\infty}^{\infty} dh e^h q_{RLG}(\Delta;h)(q_\mathrm{RLG}(\Delta;h)-1) \right )  \ ,
\end{equation}
which gives the annealed cage dimension
\begin{equation}\label{eq:phi_ann}
    \widehat{\varphi}^{-1}  = -2 \Delta \int_{-\infty}^{\infty} dh e^h (2q_\mathrm{RLG}(\Delta;h)-1) \partial_{\Delta} q_\mathrm{RLG}(\Delta;h) \ .
\end{equation}

We wish to compare the RS computation with the PHT result for $r=1$ in the limits $\widehat{\varphi}\to\infty$ and $d\to\infty$ (the only case in which the two are expected to coincide).
Note that the finite-dimensional analysis of the zero cell (cage) of the PHT with $r=1$ gives for the volume \cite[Proposition 1]{horrmann_volume_2014})
\begin{equation}\label{eq:vol_cell}
\log(\mathbb{E}[V_\mathrm{cell}])=
d\left ( -\log (d)+\log \left(\frac{\pi }{\widehat{\varphi} }\right) +\log \left(\frac{\Gamma \left(\frac{d+1}{2}\right)}{\Gamma \left(\frac{d}{2}+1\right)}\right)+\frac{\log \left(\frac{\Gamma (d+1)}{\Gamma \left(\frac{d}{2}+1\right)}\right)}{d}\right)\ ,
\end{equation}
after the change of variable $\gamma = d^2 \widehat{\varphi}/2$. We first rescale by $d$ and remove the common $\log(d)$ shift. We then take the $d\to\infty$ limit, which for Eq.~\eqref{eq:vol_cell} gives
\begin{equation}
\log(V_\mathrm{cell})/d+\log(d)/d=
-\log  \left(\widehat{\varphi} \right) + \log \left(2\pi\right)-\frac{1}{2}\ ,
\end{equation}
Although the limit $d\to\infty$ of Eq.~\eqref{eq:vol_cage} cannot be evaluated analytically for arbitrary $\widehat{\varphi}$, the expression simplifies in the limit $\widehat{\varphi}\to\infty$ ($\Delta\to 0$).
After changing variable $h\to \sqrt{\Delta} x$, we have that 
Eq.~\eqref{eq:phi} and Eq.~\eqref{eq:phi_ann} both give
\begin{equation}
    \widehat{\varphi}^{-1} \approx  K  \sqrt{\Delta}
\end{equation}
with $K_\mathrm{qch} = \frac{1}{\pi}\int_{-\infty}^{\infty} dx \frac{e^{-2x^2}}{1+\text{erf}(x)} \approx 0.638657$ for the quenched computation and $K_\mathrm{ann} = \frac{1}{\sqrt{\pi}}\int_{-\infty}^{\infty}e^{-x^2} x \, \text{erf}(x) = 1/\sqrt{2\pi}$ for the annealed one. After inserting this asymptotic behavior for large $\widehat{\varphi}$  in Eq.~\eqref{eq:vol_cage} and Eq.~\eqref{eq:vol_cage_ann}, we get
\begin{equation}
\log(V_\mathrm{cage})/d+\log(d)/d=-\log \left(\widehat{\varphi}\right)+\frac{\log (2\pi )}{2}-\log \left(K\right)  -\frac{1}{2}\ ,
\end{equation}
where we have used the identity $\int_{-\infty}^{\infty} dx \frac{1}{2} \left(\text{erf}\left(x\right)+1\right) \log \left(\frac{1}{2} \left(\text{erf}\left(x\right)+1\right)\right) = -K_\mathrm{qch} \approx -0.638657$ and $\int_{-\infty}^{\infty} dx\frac{1}{4} \left(\text{erf}(x)^2-1\right) =  -K_\mathrm{ann} = -1/\sqrt{2\pi}$.
Therefore, in the limit of high density, the annealed volume does match the PHT cell volume for $r=1$ .

\section{\lowercase{h}RLG volume in finite $d$}

\begin{figure}[t]
	\includegraphics[width=0.29\columnwidth]{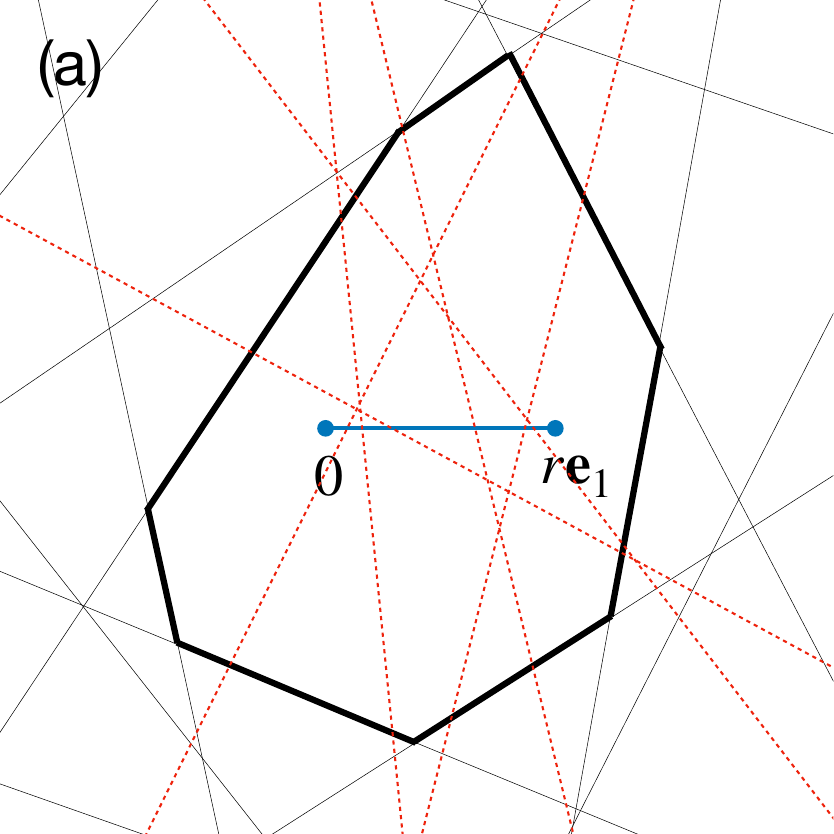}
    \includegraphics[width=0.29\columnwidth]{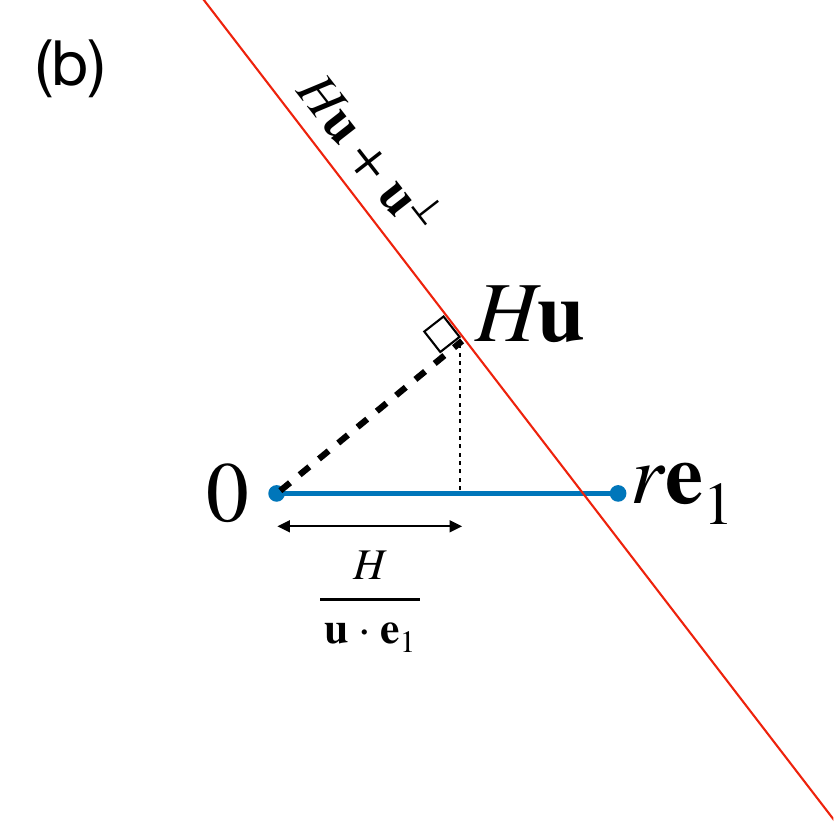}
    \includegraphics[width=0.40\columnwidth]{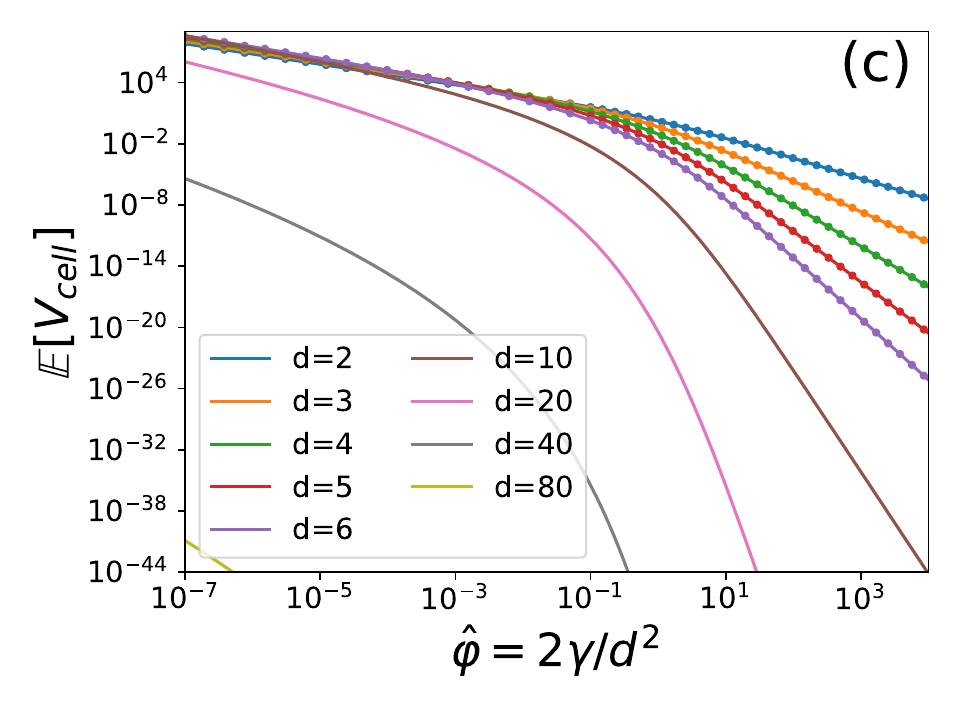}
		\caption{Setup for the stochastic geometry calculation of the average volume. (a) The ensemble of hyperplanes (black lines) that do not intersect the segment $[0,r \mathbf{e}_1]$ (blue line) as well as the forbidden hyperplanes (dashed red lines). (b) The intersection between a hyperplane (red line) and the segment $[0,r \mathbf{e}_1]$ (blue line) depends on the angle $\mathbf{u}\cdot\mathbf{e}_1$. (c) Average volume of the hRLG cell obtained from Eq.~\eqref{eq:avgV} for different $d$ (solid lines). Simulation results for $d=2,\ldots,6$ (dots) fully agree with these predictions.} \label{fig:BPlot}
\end{figure}

The average volume $\mathbb{E}[V_\mathrm{cell}]$ of the hRLG  can be exactly calculated using a standard zero cell computation from stochastic geometry. 
In order to do so, we first need to reintroduce the hRLG cell more formally. Given a Poisson hyperplane process $\eta$ with intensity measure $2\gamma \mu$ for $\mu$ the rotation invariant measure of density introduced in Eq.~\ref{eq:Ph}. The volume of the zero cell (i.e. the cell around the origin) can then be written as
\begin{equation}
\mathbb{E}[V_\mathrm{cell}] 
= \mathbb{E}[\int_{\mathbb{R}^d} d\mathbf{x} \; \mathbbm{1}(\mathbf{x}\in \mathrm{cell})]= \int_{\mathbb{R}^d} d\mathbf{x} \; \mathbb{P}(\mathbf{x}\in \mathrm{cell})= \omega_d \int_{0}^{\infty}  dr r^{d-1} \; \mathbb{P}(r \mathbf{e}_1\in \mathrm{cell})
\end{equation}
where $\omega_d = d V_d$ is the surface of the unit ball (integrated because of rotational invariance) and $\mathbf{e}_1$ is a unit vector of the canonical base of $\mathbb{R}^d$. The probability that the vector $r \mathbf{e}_1$ lies inside the zero cell is equivalent to the probability that the segment $[0,r \mathbf{e}_1]$ does not intersect any of the Poisson hyperplanes $\mathcal{H}$ (Fig.~\ref{fig:BPlot}a)
\begin{equation}
    \mathbb{P}(r \mathbf{e}_1\in \mathrm{cell}) = \mathbb{P}(\#\{\mathcal{H}\in\eta: \mathcal{H}\cap[0,r \mathbf{e}_1]\neq \emptyset \} = 0) = e^{-2\gamma \mu (\{\mathcal{H}\in\eta: \mathcal{H}\cap[0,r \mathbf{e}_1]\neq \emptyset \})},
\end{equation}
where the last expression is specific to a Poisson process.
Assuming the radial density of the hRLG $P(H)=(H+l)^{d-1}$, we have
\begin{equation}
\begin{aligned}
\mu (\{\mathcal{H}\in\eta: \mathcal{H}\cap[0,r \mathbf{e}_1]\neq \emptyset \}) 
&= \int_{S^{d-1}} \frac{\sigma(d\mathbf{u})}{\omega_d}\int_0^\infty dH \mathbbm{1}\big((H \mathbf{u}+\mathbf{u}^{\perp})\cap [0,r \mathbf{e}_1]\neq \emptyset\big)(H+l)^{d-1}\\
&=\int_{S^{d-1}} \frac{\sigma(d\mathbf{u})}{\omega_d}\int_0^\infty dH \mathbbm{1}\big(\frac{H}{\mathbf{u}\cdot\mathbf{e}_1}\in [0,r]\big)(H+l)^{d-1}
\end{aligned}
\end{equation}
where $S^{d-1}$ is the surface of the $d$-dimensional ball and $\sigma(d\mathbf{u})$ is the spherical Lebesgue measure. Note that $\mathbf{u}$ is a unit vector and $\mathbf{u}^{\perp}$ is a base perpendicular to it (that spans the hyperplane) (see Fig.~\ref{fig:BPlot}b). Because $\mathbf{u}$ is spherically uniform, the density can be expressed in terms of the scalar product $u_1\equiv\mathbf{u}\cdot\mathbf{e}_1$ alone, see \cite[Lemma 4.4. (a)]{kabluchko2019expected} for example.
Therefore,
\begin{equation}
\begin{aligned}
\mu(\{\mathcal{H}\in\eta: \mathcal{H}\cap[0,r \mathbf{e}_1]\neq \emptyset \})
&=\int_{-1}^{1}du_1 c_d(1-u^2_1)^\frac{d-3}{2}  \mathbbm{1}(u_1\ge 0) \int_0^{r u_1} dH (H+l)^{d-1}\\
&= \frac{c_d}{d}\int_{0}^{1}du_1 (1-u^2_1)^\frac{d-3}{2} \big((l+ru_1)^d-l^d\big) 
\end{aligned}
\end{equation}
where $c_d=1/\int_{-1}^{1}du_1 (1-u^2_1)^\frac{d-3}{2} = \Gamma(\frac{d}{2})/  \Gamma(\frac{d-1}{2}) \sqrt{\pi} $ is a normalization factor and the last identity is obtained by elementary manipulations. Therefore the average volume of a hRLG cell is 
\begin{equation}
\label{eq:avgV}
\mathbb{E}[V_\mathrm{cell}] 
=  \omega_d \int_{0}^{\infty}  dr r^{d-1} e^{-2\gamma \frac{c_d}{d} \int_{0}^{1}du_1 (1-u^2_1)^\frac{d-3}{2}\big((l+ru_1)^d-l^d\big) }
\end{equation}
Finally, to obtain the volume as a function of the packing fraction we substitute $\gamma \to \frac{d^2\widehat{\varphi}}{2}$.
Evaluating Eq.~\eqref{eq:avgV} numerically agrees closely with  simulations results (see Fig.~\ref{fig:BPlot}c). Note that it is trivial to generalize the calculation to arbitrary radial density $P(H)$ with corresponding cumulative distribution function $C(H)=\int_{0}^{H}dH' \,P(H')$. It indeed suffices to substitute $\big((l+ru_1)^d-l^d\big)/d$ by $C(ru_1)$ in the above treatment.

%% file: main.bbl
\begin{thebibliography}{58}
\bibitem [{mat()}]{math}%
  \BibitemOpen
  \href@noop {} {}\bibinfo {note} {In the mathematical literature the volume
  fraction is replaced by the intensity $\gamma = \frac{d^2
  \widehat{\varphi}}{2}$. See \cite[Eq.~(3)]{horrmann_volume_2014} for a more
  formal definition.}\BibitemShut {Stop}%
\bibitem [{Ph()}]{Ph}%
  \BibitemOpen
  \href@noop {} {}\bibinfo {note} {The first option is asymmetric in that the
  central particle has its own potential and the obstacles are hard with
  well-defined tangent hyperplanes. The second option is symmetric in that both
  the central particle and the obstacles interact through the same potential.
  The tangent planes are then soft, and hence the PHT framework is no longer
  valid, but the mean-field $d\to\infty$ one remains.}\BibitemShut {Stop}%
\end{thebibliography}
